\documentclass[letterpaper, 10 pt, conference]{ieeeconf}
\IEEEoverridecommandlockouts
\usepackage{chngpage}
\usepackage{tabularx}
\usepackage{graphicx}
\usepackage{siunitx}
\usepackage{hyperref}
\hypersetup{colorlinks=true, citecolor=blue, linkcolor=blue, pdfpagemode=UseNone}

\usepackage{graphics, multirow, booktabs}
\usepackage{amsmath, amssymb}

\let\labelindent\relax
\usepackage[inline]{enumitem}

\title{\LARGE \bf
Mental arithmetic task classification with convolutional neural network based on spectral-temporal features from EEG
}

\author{Zaineb Ajra, Binbin Xu, Gérard Dray, Jacky Montmain, Stephane Perrey
\thanks{$^{1}$EuroMov Digital Health in Motion, Univ. Montpellier, IMT Mines Ales, Ales, France. 
        {{\tt\small zaineb.ajra@umontpellier.fr}
        \tt\small binbin.xu@mines-ales.fr} \newline 
		DOI: \href{https://doi.org/10.1109/EMBC48229.2022.9870887}{10.1109/EMBC48229.2022.9870887} 
        }%
}

\begin{document}

\maketitle
\setcounter{page}{52}
\renewcommand{\thepage}{\bf\arabic{page}}
\thispagestyle{plain}
\pagestyle{plain}

\begin{abstract}
In recent years, neuroscientists have been interested to the development of brain-computer interface (BCI) devices. Patients with motor disorders may benefit from BCIs as a means of communication and for the restoration of motor functions. Electroencephalography (EEG) is one of most used for evaluating the neuronal activity.
In many computer vision applications, deep neural networks (DNN) show significant advantages.
Towards to ultimate usage of DNN, we present here a shallow neural network that uses mainly two convolutional neural network (CNN) layers, with relatively few parameters and fast to learn spectral-temporal features from EEG. We compared this models to three other neural network models with different depths applied to a mental arithmetic task using eye-closed state adapted for patients suffering from motor disorders and a decline in visual functions. 
Experimental results showed that the shallow CNN model outperformed all the other models and achieved the highest classification accuracy of $90.68\%$. It's also more robust to deal with cross-subject classification issues: only $3\%$ standard deviation of accuracy instead of $15.6\%$ from conventional method. 

\end{abstract}

\section{INTRODUCTION}

Nowadays, one of the medical challenges is to understand the diseased brain in the context of neurological diseases such as Alzheimer's, Parkinson's, consciousness disorders, epilepsy, etc., in order to develop suitable treatments. The brain and the spinal cord constitute the central nervous system, capable of integrating information, controlling motor skills / movements and ensuring cognitive functions \cite{ito2002controller}. Several technologies based on head movements, eye contact, etc have given people the chance to control a wheelchair or robot. However, for patients with brain injury who have lost muscle control voluntarily, it is not easy or even not possible to use these technologies.

The brain computer interface (BCI) technology offers those who are deprived of muscle control a way to communicate with their environment and become less dependent in their daily activities \cite{ma2021fnirs}\cite{khan2017hybrid}. 
One of the most common BCI is through electroencephalography (EEG), a non-invasive neuroimaging technique that measures brain activity, which is more accessible  and more accurate than functional magnetic resonance imaging (fMRI) \cite{asgher2020classification} in time resolution. 
EEG signals reflect valuable information about the brain activity in measured scalp regions. 
This is crucial for the diagnosis and management of certain diseases or brain injuries. 

In consequence, the first step is to better classify the acquired EEG signals, for example separating EEG signals containing cognitive activities from the resting state. 
However, classifying cognitive tasks using EEG signals necessitates solving difficult pattern recognition issues. A prior feature engineering step is required with conventional classification methods. Due to the highly complex property of EEG signal, these methods cannot always guarantee to produce the acceptable performance of classification. The neural network (NN) in computer science has reached human parity milestone in many applications. The most exciting feature of neural network is its self-learning ability which implies that manual feature extraction is not necessarily required. This approach is now more and more used in EEG classification to overcome the limits of conventional methods, especially when data volume is increasing in an exponential way  \cite{ghonchi2020deep}\cite{saadati2019mental}.

One main issue when using NN is the choice of feature selection and architecture, as in other fields. 
In this work, we proposed to use joint spectral-temporal features instead of the ones extracted with common spatial pattern  filtering and to benchmark different NN architectures for a mental task classification using EEG signals, from basic shallow NN to state-of-art deep NN models.

\section{Data and Models}

\subsection{Data from cognitive tasks}

\subsubsection{Participants and data}
In this study, we used the publicly available dataset collected by Shin et al. in 2018 \cite{shin2018eyes}. In the original dataset, 22 EEG channels and 9 near-infrared spectroscopy (NIRS) channels were simultaneously recorded from the scalp in 12 participants performing mental arithmetic (MA) task and being relaxed (BL). Here, we only consider EEG recordings. 
Regarding the EEG signals acquired, 10 were placed on the frontal cortex and 12 other on parieto-occipital areas.

\subsubsection{Experimental protocol}
Experiments were conducted in 3 sessions. In each session, a pre-rest period of \SI{15}{\second} was followed by 20 trials of task phase (27--\SI{29}{\second} each) and then a final \SI{15}{\second} post-rest period. A fixation cross was shown on the monitor through the pre-rest and post-rest periods. During each trial, participants were instructed to look at a visual instruction (\SI{2}{\second}) indicating the type of task, followed by task period of \SI{10}{\second} and a rest period (15--\SI{17}{\second}). In the instruction period, the type of task was randomly displayed on the screen (MA or BL). For MA task, an arbitrary one-digit number between 6 and 9 was subtracted from three digit number. For BL, participants were asked to relax and gaze at the fixation cross displayed in the middle of the monitor. A total of 60 trials (20 trials $\times$ 3 sessions) were performed  by each subject.

\subsubsection{Data preprocessing}
The raw EEG signals from 22 channels were down-sampled to \SI{200}{\hertz} and then band-pass filtered within the range of 0.5--\SI{50}{\hertz} using a 3rd-order Butterworth filter. Ocular artifacts were rejected manually. The artifact-free data was segmented into epochs using time intervals from \SI{-2}{\second} (\SI{2}{\second} before the beginning of the task) to \SI{10}{\second} (the end of the task). The baseline in the range of \SI{-1}{\second} to \SI{0}{\second} were removed from data. In total, 60 epochs of \SI{12}{\second} duration were obtained for each participant.  


\begin{figure}[!htb]
\centering
  \includegraphics[width=\columnwidth]{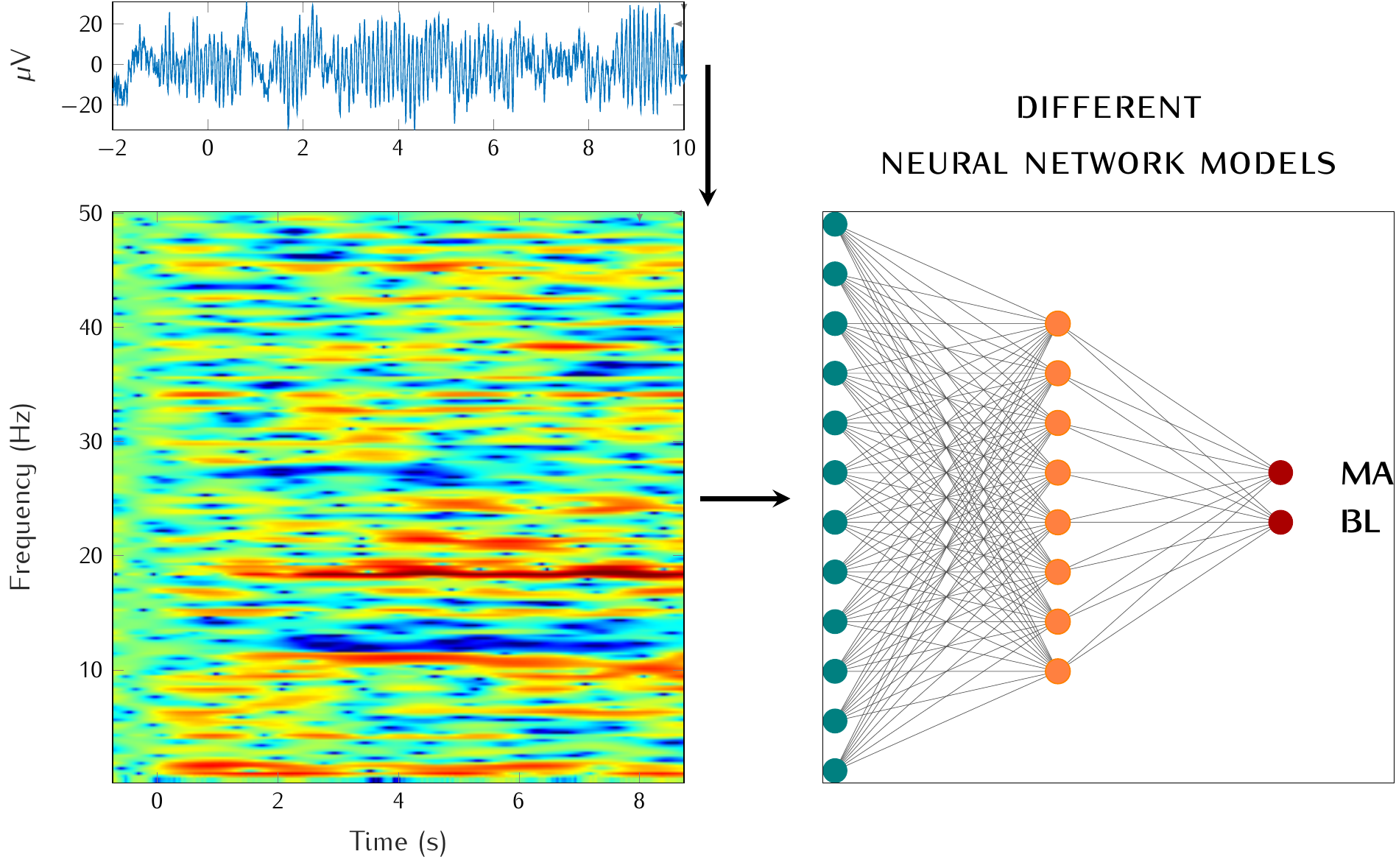}
  \caption{The study work-flow: from signal-spectrum to different models}
  \label{fig:workflow}
\end{figure}

\subsection{Spectral-temporal feature extraction}
In the work \cite{shin2018eyes}, Shin et al. used features extracted from EEG signals with common spatial pattern filter (CSP). The time-frequency analysis allows to obtain richer information which are more appropriate to neural network. Thus, in this work, spectral-temporal features were used. These features are extracted with event-related spectral power (ERSP) approach from EEG data with EEGLab. ERSP is applied at epoch level. As most known DNN have fixed input dimensions such as ($224 \times 224$), to take the advantage of these powerful models, the ERSP output (time $\times$ frequency domain) is set with the same dimensions. 
In total $15\,480$ spectrograms (12 subjects $\times$ 22 electrodes $\times$ 60 epochs) were extracted. 
The main workflow can be found in Fig. \ref{fig:workflow}.

\subsection{Models}

Shin et al. used shrinkage linear discriminant analysis (sLDA) to classify the features obtained from CSP on EEG. While classifying MA task from BL, the performance by sLDA reached $77.3 \pm 15.9\%$ (mean $\pm$ std) in terms of accuracy. 
In our work, we opted for neural network models. Two types of NN models were evaluated:
\begin{enumerate*}
    \item proposed shallow neural network: basic recurrent neural network ({RNN}) and convolutional neural network;
    \item common state-of-art deep neural networks: GoogLeNet and ResNet-50.
\end{enumerate*}

\subsubsection{Long Short-Term Memory (LSTM) model}

\begin{figure}[!htb]
\centering
  \includegraphics[width=.8\columnwidth]{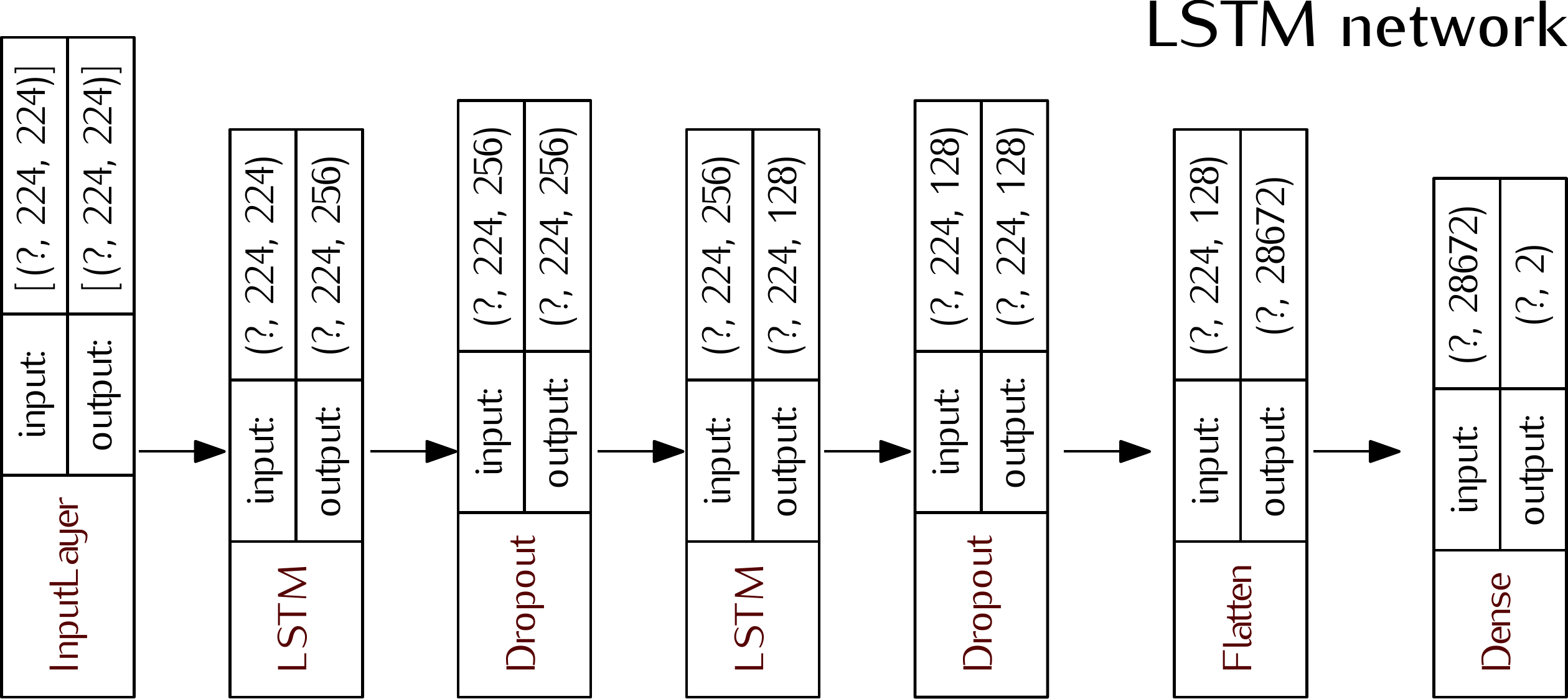}
  \caption{Baseline LSTM model architecture}
  \label{fig:model_lstm}
\end{figure}

RNN is capable to learn temporal dependence in the sequential data. It's the common architecture to deal with time series data, and often serves as the baseline model in bench-marking. Here, we proposed to use a first model based on LSTM which is one of the most used RNN variants. This model is consisted of two LSTM layers having respectively $256$ and $128$ hidden neurons. Each of them is followed by a Dropout layer (drop rate 0.5). The output classification layer used {\sc{softmax}} as activation. The spectral-temporal features are thus considered as for example $224$ time series in the frequency domain. The model architecture can be found in Fig. \ref{fig:model_lstm}.

\subsubsection{Convolutional Neural Networks}

CNN is currently the most used neural network architecture. The main advantage of CNN is its self-learning ability of feature extraction -- using a series of convolution filters, CNN layer can generate invariant features from two-dimensional / three-dimensional data. There's no necessarily further step of manual feature extraction. 
Another advantage of CNN models is that it also takes the spatial (or cross-dimensional in case of data other than images) constraints which may yield better features. 
The two-dimensional spectral-temporal features ERSP features of dimension $224 \times 224$ from EEG signals can be considered as one type of image. Application of CNN layers on its classification can then be considered. 

\begin{figure}[!htb]
\centering
\includegraphics[width=0.8\columnwidth]{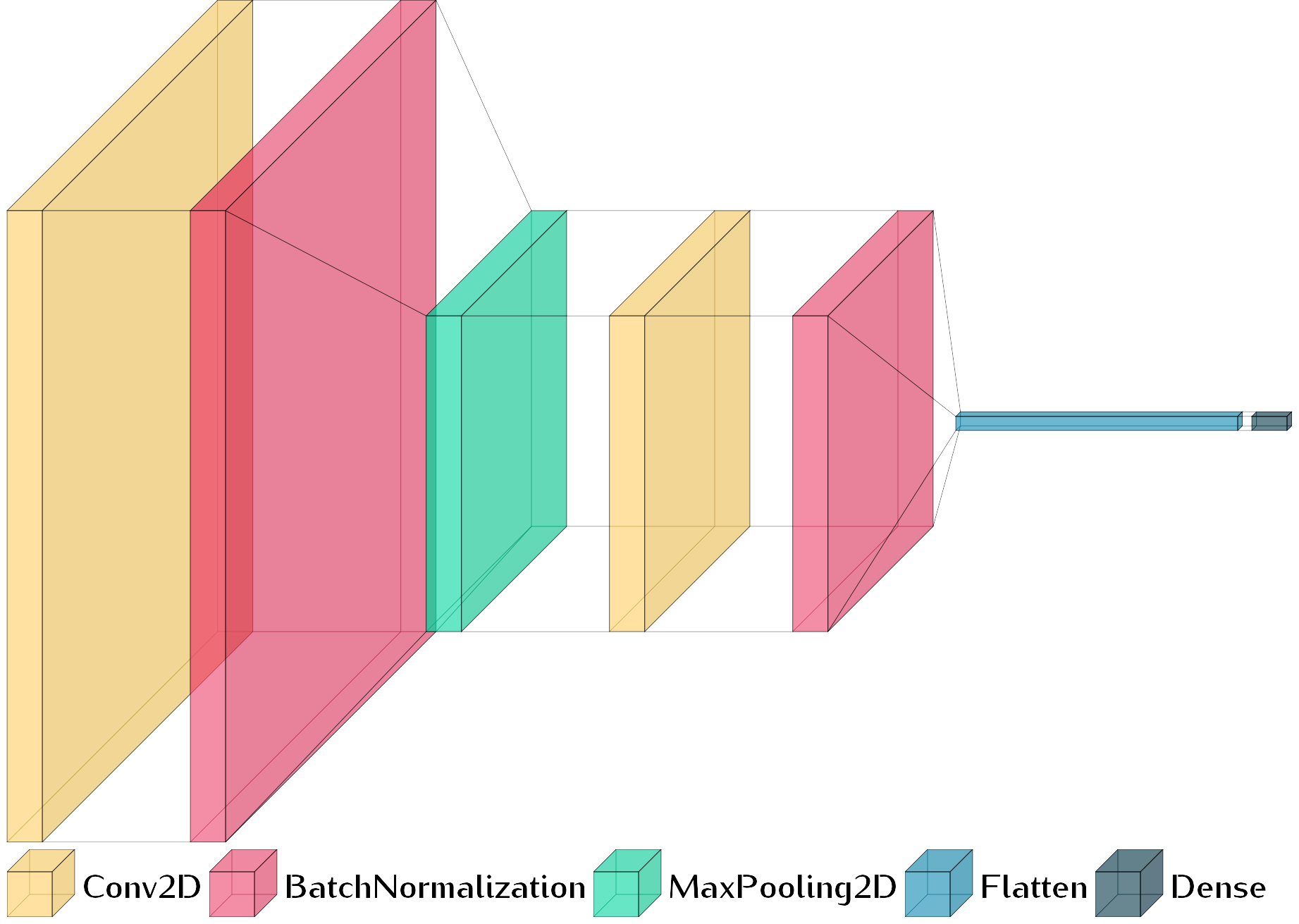}
\caption{CNN model architecture}
\label{fig:model_cnn}
\end{figure}

The proposed CNN model (Fig. \ref{fig:model_cnn}) followed the same principle -- shallow neural network. This model has essentially two 2D convolutional layers (Conv2D) which are each attached to a batch normalization layer to address the issue of internal covariate shift. The Batch normalization layer can also be considered as regularizer, in many cases it may help to eliminate the need of a Dropout layer. 
Both Conv2D layers using a Rectified Linear Unit (ReLU) activation have 10 filters with kernel of size $3\times3$ and zero-padding is applied to match the same input dimension. 
These two blocks are linked by a max pooling layer with $2\times2$ size and $2\times2$ stride.

\subsubsection{Deep neural network}

Considering the data size limit, two proposed models are both basic shallow network. The ultimate goal is to acquire enough data to re-train the stat-of-the-art DNN models. At the first stage, it's worth to investigate how these DNN models perform in this application. 
Two models with the same input image size are considered: GoogLeNet and ResNet-50. As shown in Table \ref{tab:nn_property}, they have considerably larger sizes in comparison to the proposed shallow networks which have only $0.74$ / $0.25$ million parameters. 

\begin{table}[htbp]
  \centering
  \caption{Models properties}
    \begin{tabular}{lccc}
    \toprule
    \textbf{Network} & {\textbf{Depth}} & {\textbf{Parameters (Millions)}} & \textbf{Input Size} \\
    \midrule
    GoogLeNet & 22    & 7     & 224-by-224 \\
    ResNet-50 & 50    & 25.6  & 224-by-224 \\
    \midrule
    LSTM & 2 & 0.74 & 224-by-224 \\
    CNN  & 2 & 0.25 & 224-by-224\\
    \bottomrule
    \end{tabular}%
  \label{tab:nn_property}%
\end{table}%

\subsection{Training options and performance evaluation}

All models were trained in the same configuration. Stochastic gradient descent with momentum (SGDM) optimizer is used for all. The initial learning rate is fixed at $0.001$. The maximum number of training epoch is 50, with batch size 64. The same early stopping rule is applied -- validation patience of 20 and a validation frequency of per 8 iterations. 

Keeping the original subject / task ratios from the raw data, the $15\,840$ samples are randomly split into {\sc training} ($70\%$), {\sc validation} ($15\%$) and {\sc test} ($15\%$). 
In order to minimize the risk of over-fitting, 20 random triple-sets were generated and used for all models. 
The median values of commonly used metrics (Accuracy, Specificity, Sensitivity and F1-score) on {\sc test} sets are reported from the 20 splits. 


\begin{table}[!b]
\setlength{\tabcolsep}{5pt}
  \centering
  \caption{The performances of different methods (median $\%$)} 
    \begin{tabular}{lcccc}
    \toprule
    \textbf{Methods } & \textbf{ACC} & \textbf{Sensibility} & \textbf{Specificity} & \textbf{F1 score} \\
    \midrule
    sLDA \cite{shin2018eyes} & $80.10$  & --  & --  & --\\ 
    \midrule
    LSTM & $67.76$ & $67.74$ & $68.37$  & $68.38$\\
    GoogLeNet & $73.34$ & $77.56$  & $72.45$ & $72.22$ \\
    ResNet-50 & $72.77$ & $74.63$ & $72.00$  & $72.07$\\
    \midrule
    CNN & $\mathbf{90.68}$ & $\mathbf{94.34}$ &  $\mathbf{90.85}$ & $\mathbf{90.21}$\\ 
    \midrule
    \\
    \multicolumn{5}{c}{\sc{standard deviation of CNN model }} \\
    \midrule
    CNN \emph{std}      & $\mathbf{2.30}$  & $\mathbf{5.53}$ &  $\mathbf{6.29}$  & $\mathbf{2.28}$ \\ 
    \bottomrule
    \end{tabular}%
  \label{tab:model_perf}%
\end{table}%

\section{Results / Discussion}

For the four NN models with same training conditions -- shallow LSTM / CNN, deep GoogLeNet and ResNet-50, their performance metrics are presented in Table \ref{tab:model_perf}. 
The shallow LSTM model showed poor performance, with a median accuracy only at $67.76\%$ which is quite lower than the baseline results of $80.1\%$ from \cite{shin2018eyes}.
Deep NN models performed similarly. Both models  showed neither satisfactory results with $73.34\%$ / $72.77\%$ accuracy. 
The shallow CNN model, despite its smallest number of parameters, achieved an accuracy of $90.68\%$ and outperformed all the listed models. This model is also much more robust with a standard deviation accuracy only at $2.3\%$, while it was $15.9\%$ in \cite{shin2018eyes}. 

\begin{table}[!htbp]
  \centering
  \caption{Performances by Participant with CNN model (median \%)}
    \begin{tabular}{c|c|cccc}
    \toprule
    & \textbf{sLDA \cite{shin2018eyes}} &\multicolumn{4}{c}{\textbf{proposed CNN model}} \\
    \midrule
    \textbf{Participant} & \textbf{ACC} {\tiny{EEG}} & \textbf{ACC} & \textbf{Sen} & \textbf{Spe} & \textbf{F1} \\
    \midrule
	 1 & $96.5$ & $90.3 \pm 4.4$ & 	$92.7$ & 	$91.3$ & 	$90.1$ \\
	 2 & $79.0$ & $91.2 \pm 3.0$ & 	$94.4$ & 	$93.4$ & 	$91.3$ \\
	 3 & $58.2$ & $92.0 \pm 2.6$ & 	$94.6$ & 	$94.0$ & 	$91.8$ \\
	 4 & $90.7$ & $90.9 \pm 2.9$ & 	$93.7$ & 	$91.2$ & 	$90.9$ \\
	 5 & $95.7$ & $90.3 \pm 3.4$ & 	$91.1$ & 	$91.6$ & 	$89.9$ \\
	 6 & $83.0$ & $89.4 \pm 3.6$ & 	$91.6$ & 	$89.1$ & 	$89.6$ \\
	 7 & $50.7$ & $92.1 \pm 2.6$ & 	$95.1$ & 	$90.8$ & 	$92.1$ \\
	 8 & $66.2$ & $90.3 \pm 2.8$ & 	$92.2$ & 	$90.6$ & 	$89.9$ \\
	 9 & $81.2$ & $88.7 \pm 3.0$ & 	$91.7$ & 	$90.6$ & 	$88.3$ \\
	10 & $76.8$ & $90.1 \pm 2.7$ & 	$95.1$ & 	$91.3$ & 	$89.8$ \\
	11 & $96.0$ & $92.4 \pm 2.4$ & 	$95.7$ & 	$92.5$ & 	$92.2$ \\
	12 & $53.7$ & $89.1 \pm 2.6$ & 	$90.5$ & 	$90.1$ & 	$88.6$ \\
    \midrule
    std & $15.9$ & $1.20$ &	$1.80$  & $1.36$ & $1.30$ \\
    \bottomrule
    \end{tabular}%
  \label{tab:perf_subj}%
\end{table}%

Another common issue of classification study in neuroscience is that the performance varies strongly from one subject to another. So, we tested the trained models for each participant in test set as well. 
As shown in Table \ref{tab:perf_subj}, the CNN model demonstrated its robustness. The classification accuracy is quite similar for all the 12 participants. 
The sLDA approach with CSP extracted features failed to classify the MA task from BL, with accuracy close to $50\%$ -- the random selection in case of binary classification. The inter-participant standard deviation of accuracy is $15.6\%$, while the new CNN model showed significantly small as $1.2\%$. At participant level, the inter-participant std values are all close to $3\%$. 

\renewcommand*{\thefootnote}{\fnsymbol{footnote}}
\begin{figure}[!htb]
\centering
    \includegraphics[width=.95\columnwidth]{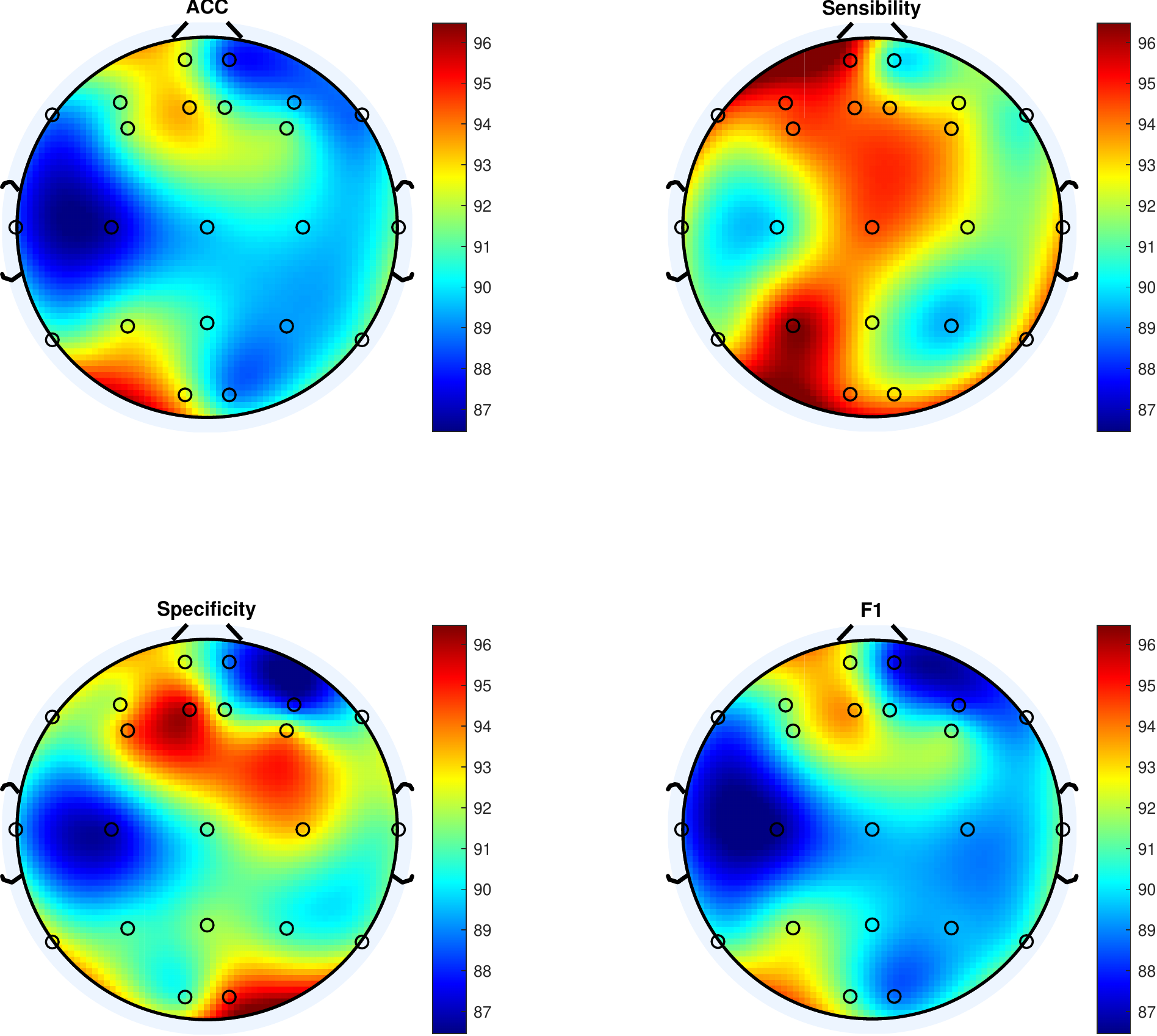}
    \caption{Topological representation of performances by Channels. {\protect\footnotemark{}}}
    \label{fig:perf_channel}
\end{figure}

Further observation at EEG channel level revealed that the performance varied from one brain region to another (Fig. \ref{fig:perf_channel}, topological plot with interpolation from EEGLab; detailed values in Tab. \ref{tab:perf_channel}). 
The good accuracy in the frontal cortex region (especially left) is consistent with literature on mental arithmetic tasks \cite{sammer2007relationship}. 
In occipital region, similar accuracy was obtained. This suggested that this region might be a promising candidate in the classification of this type of tasks. 

\begin{table}[htbp]
  \centering
  \caption{Performances by Channels (median \%)}
    \begin{tabular}{lc|lc|lc}
    \toprule
    \textbf{Channel} & {\textbf{ACC}} & \textbf{Channel} & \textbf{ACC} & \textbf{Channel} & \textbf{ACC} \\
    \midrule
    F7    & $89.1$  & AFF1h & $93.3$  & POO1 & $92.6$ \\
    AFF5h & $91.3$  & AFF2h & $91.3$  & POO2 & $89.1$ \\
    F3    & $91.7$  & Cz    & $89.7$  & P4   & $89.2$ \\
    AFp1  & $92.2$  & Pz    & $90.5$  & P8   & $91.4$ \\
    AFp2  & $88.1$  & T7    & $88.5$  & C4   & $90.0$ \\
    AFF6h & $89.3$  & C3    & $86.9$  & T8   & $91.4$ \\
    F4    & $91.4$  & P7    & $91.5$  &      &  \\
    F8    & $88.8$  & P3    & $92.3$  &      &  \\
    \bottomrule
    \end{tabular}%
  \label{tab:perf_channel}%
\end{table}

The poor performance from LSTM models is generally related to one of its drawbacks - lack of spatial (or cross-dimensional) learning. More specifically in this study, LSTM model took only the temporal information into account, the structural correlation in frequency domain being ignored. This would make it fail to classify correctly this type of data in cognitive task. 
Applying the same early stopping rules as shallow NN, the deep neural networks showed only moderate performance. However, if they are trained differently (with less strict early stopping rule and considerably more fine-tuning), these models started to show promising performance (some case better than baseline result with sLDA). They should perform better when the data size is sufficiently large.
\begin{center}
  \footnotesize
    \begin{tabular}{lcccc}
    \multicolumn{5}{c}{\sc{Further training on LSTM and ResNet-50 models}} \\
    \toprule
    \textbf{Methods } & \textbf{ACC} & \textbf{Sensibility} & \textbf{Specificity} & \textbf{F1 score} \\
    \midrule
    LSTM & $78.87$ & $82.69$ &  $79.01$  & $75.86$\\ 
    ResNet-50 & $85.90$ & $88.92$ & $83.48$  & $84.86$\\
    \bottomrule
    \end{tabular}%
\end{center}%
But one needs to note that the training time is significantly longer than shallow networks. And many deep NN models run only on graphics processing unit (GPU) in order to be trained in an acceptable time range. 
When the data size is limited, this study highlights that use of shallow CNN may already help to obtain satisfactory accuracy as high as $91\%$. There exists still slight room to improve the performance of this model by fine-tuning the architecture and training parameters. 

\footnotetext{\emph{Figure 4 uploaded to IEEE Explore was not compiled correctly. This is the updated one.}}

\section{Conclusion}
In this study, we examined the performance of four neural network models in cognitive tasks classification of EEG recordings at trial level. Our findings revealed that the proposed shallow CNN improves significantly classification performance reaching $90.68\%$ cross-validation accuracy, which is $10.6\%$ higher than with sLDA method in their original study \cite{shin2018eyes}. 
The spectral and temporal features from EEG would be more representative than features extracted by spatial pattern filtering. 
In comparison with CNN model, the basic LSTM model in time domain only showed poor performance and failed to accomplish the classification task. 
This suggests that the structure across frequencies can indeed bring additional information.

Even though the DNN did not give satisfactory results when using the same training options, when trained with different configurations, their performances started to increase and can surpass the performance with sLDA. However, this will require much more training time and parameter fine-tuning. They are still very promising. Increasing the data volume will help to improve the performance and most importantly to propose some more generalized and robust models when dealing with unlearned new data. A project of this objective is currently on-going.

The high performance of the shallow CNN model showed that for cognitive task classification, simple shallow network might already be sufficient which requires much less training time and computational load. The personalised model is also conceivable when trial-level data number is enough. This can also help to reduce the inter-subjects issue. 

\section*{ACKNOWLEDGMENT}
Zaineb AJRA received a doctoral fellowship from \mbox{AXIAUM} Univ. Montpellier-ISDM ({\small ANR-20-THIA-0005-01}) and ED I2S in France.

\bibliographystyle{IEEEtran}
\bibliography{bib}

\begin{thebibliography}{1}
\providecommand{\url}[1]{#1}
\csname url@samestyle\endcsname
\providecommand{\newblock}{\relax}
\providecommand{\bibinfo}[2]{#2}
\providecommand{\BIBentrySTDinterwordspacing}{\spaceskip=0pt\relax}
\providecommand{\BIBentryALTinterwordstretchfactor}{4}
\providecommand{\BIBentryALTinterwordspacing}{\spaceskip=\fontdimen2\font plus
\BIBentryALTinterwordstretchfactor\fontdimen3\font minus
  \fontdimen4\font\relax}
\providecommand{\BIBforeignlanguage}[2]{{%
\expandafter\ifx\csname l@#1\endcsname\relax
\typeout{** WARNING: IEEEtran.bst: No hyphenation pattern has been}%
\typeout{** loaded for the language `#1'. Using the pattern for}%
\typeout{** the default language instead.}%
\else
\language=\csname l@#1\endcsname
\fi
#2}}
\providecommand{\BIBdecl}{\relax}
\BIBdecl

\bibitem{ito2002controller}
M.~Ito, ``Controller-regulator model of the central nervous system,''
  \emph{Journal of integrative neuroscience}, vol.~1, no.~02, pp. 129--143,
  2002.

\bibitem{ma2021fnirs}
T.~Ma, W.~Chen, X.~Li, Y.~Xia, X.~Zhu, and S.~He, ``fnirs signal classification
  based on deep learning in rock-paper-scissors imagery task,'' \emph{Applied
  Sciences}, vol.~11, no.~11, p. 4922, 2021.

\bibitem{khan2017hybrid}
M.~J. Khan and K.-S. Hong, ``Hybrid eeg--fnirs-based eight-command decoding for
  bci: application to quadcopter control,'' \emph{Frontiers in neurorobotics},
  vol.~11, p.~6, 2017.

\bibitem{asgher2020classification}
U.~Asgher, K.~Khalil, Y.~Ayaz, R.~Ahmad, and M.~J. Khan, ``Classification of
  mental workload (mwl) using support vector machines (svm) and convolutional
  neural networks (cnn),'' in \emph{2020 3rd International Conference on
  Computing, Mathematics and Engineering Technologies (iCoMET)}.\hskip 1em plus
  0.5em minus 0.4em\relax IEEE, 2020, pp. 1--6.

\bibitem{ghonchi2020deep}
H.~Ghonchi, M.~Fateh, V.~Abolghasemi, S.~Ferdowsi, and M.~Rezvani, ``Deep
  recurrent--convolutional neural network for classification of simultaneous
  eeg--fnirs signals,'' \emph{IET Signal Processing}, vol.~14, no.~3, pp.
  142--153, 2020.

\bibitem{saadati2019mental}
M.~Saadati, J.~Nelson, and H.~Ayaz, ``Mental workload classification from
  spatial representation of fnirs recordings using convolutional neural
  networks,'' in \emph{2019 IEEE 29th International Workshop on Machine
  Learning for Signal Processing}.\hskip 1em plus 0.5em minus 0.4em\relax IEEE,
  2019, pp. 1--6.

\bibitem{shin2018eyes}
J.~Shin, K.-R. M{\"u}ller, and H.-J. Hwang, ``Eyes-closed hybrid brain-computer
  interface employing frontal brain activation,'' \emph{PloS one}, vol.~13,
  no.~5, p. e0196359, 2018.

\bibitem{sammer2007relationship}
G.~Sammer, C.~Blecker, H.~Gebhardt, M.~Bischoff, R.~Stark, K.~Morgen, and
  D.~Vaitl, ``Relationship between regional hemodynamic activity and
  simultaneously recorded eeg-theta associated with mental arithmetic-induced
  workload,'' \emph{Human brain mapping}, vol.~28, no.~8, pp. 793--803, 2007.

\end{thebibliography}

\end{document}